\documentclass[aps,prb,twocolumn]{revtex4}
\usepackage{amsmath,amssymb,bm}
\usepackage{graphicx}
\usepackage{color}

\begin{document}

\title{Decay of a plasmon into neutral modes in a carbon nanotube}
\author{Wei Chen}
\affiliation{Department of Physics, University of Washington, Seattle, WA
98195-1560, USA}

\author{A. V. Andreev}
\affiliation{Department of Physics, University of Washington, Seattle, WA
98195-1560, USA}

\author{E. G. Mishchenko}
\affiliation{Department of Physics and Astronomy, University of Utah, Salt Lake City, Utah 84112, USA}

\author{L. I. Glazman}
\affiliation{Department of Physics, Yale University, New Haven, CT 06520, USA}

\begin{abstract}
We evaluate the rate of energy loss of a plasmon in a disorder-free
carbon nanotube. The plasmon decays into neutral bosonic excitations of the electron liquid. The process is mediated either by phonon-assisted backscattering of a single
electron or Umklapp backscattering of two electrons. To lowest order in the backscattering interactions the partial decay rates are additive.
At zero doping the corresponding decay rates scale as power-laws of the temperature with positive and negative exponents for the two mechanisms, respectively. The precise values
of the exponents depend on the Luttinger liquid parameter. At finite doping the decay rates are described by universal crossover functions of frequency and chemical potential measured in units of temperature. In the evaluation of the
plasmon decay, we concentrate on a finite-length geometry allowing
excitation of plasma resonances.
\end{abstract}

\maketitle

\section{Introduction \label{sec:intro}}

Electronic properties of single wall carbon nanotubes (CNTs) have been studied
intensively in the DC limit by means of electron transport~\cite{Purewal}, by optical excitations~\cite{Bachilo,Lefebvre,Hagen,Korovyanko,Avouris}, and by angle-resolved electron energy loss spectroscopy in the UV region~\cite{Pichler2008}. CNT
properties in the intermediate (THz) band of frequencies are harder to
access experimentally and consequently they attracted less attention in
theory. The dominant excitations in that frequency range are
plasmons. Initial experiments probing them by means of
time-resolved measurements were reported recently~\cite{mceuen}. Here
we are interested in the intrinsic mechanisms of the decay of plasmons
into other excitations within a metallic (armchair) CNT.

The existing tunneling experiments provide support for the Luttinger
liquid description of the charge carriers in
CNTs~\cite{McEuen1,Dekker}. In that description, the collective
excitation, plasmon, is just one of the eigenmodes with linear
spectrum. As such, it does not decay.

In the conventional Fermi liquid description applicable to higher-dimensional conductors,
separation of excitations into plasmons and Fermi-quasiparticles is not exact,
leading to a plasmon decay by exciting quasiparticles~\cite{Mahanbook}.
In contrast, in one-dimensional conductors quasiparticles are ill-defined, and plasmon decays into other collective excitations of the electron liquid.

\begin{figure}[ptb]
\includegraphics[width=8.cm]{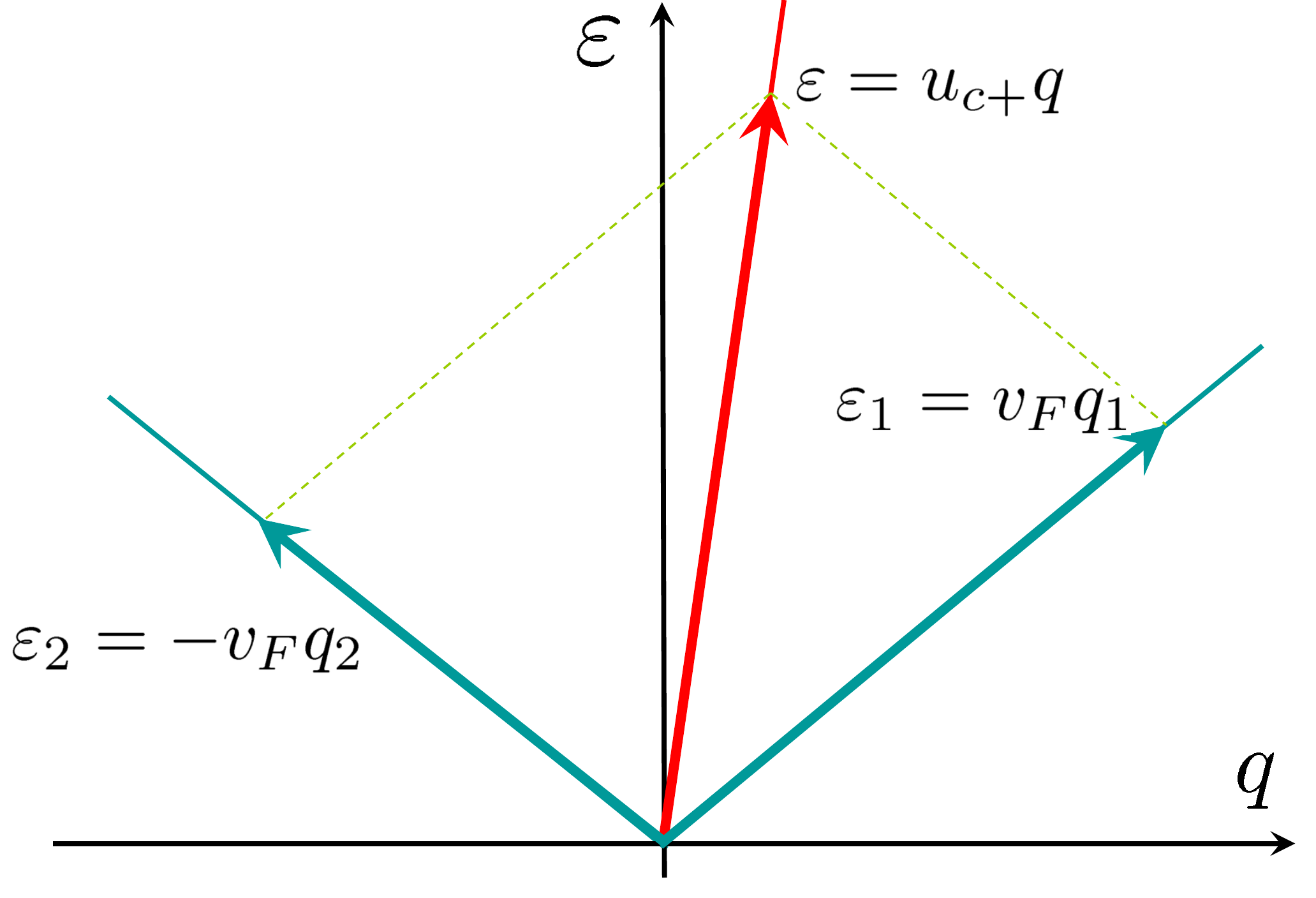}
\caption{(Color online) Schematic picture of a possible decay process
  of a plasmon excitation with momentum $q$ and energy $\varepsilon= u_{c+}q$ into two  neutral mode excitations with momenta $q_1$ and $q_2$ and energies $\varepsilon_1=v_F q_1$ and $\varepsilon_2 =-v_F q_2$. The line with the steep slope represents the plasmon
  mode with velocity $u_{c+}$ and the two lines with shallower slopes represent neutral modes
  with velocities $\pm v_F$. Conservation of energy and momentum requires $\varepsilon=\varepsilon_1+ \varepsilon_2$ and $q=q_1 +q_2$.}
\label{fig:decay}
\end{figure}

In armchair nanotubes in addition to the plasmon mode the spectrum of
bosonic low energy excitations of electron liquid contains three
neutral modes propagating with the Fermi velocity. Conservation of
energy and momentum allows a plasmon decay into neutral modes, as
illustrated in Fig.~\ref{fig:decay}. However absence of coupling
between the bosonic modes in the Luttinger liquid description
precludes the possibility of plasmon decay. It  arises
from backscattering interactions and thus requires consideration of
corrections to the Luttinger liquid approximation.

Another qualitative difference between plasmon decay in nanotubes from
that in higher dimensions is that in one dimension the
excitations of neutral modes generated during the plasmon decay can propagate only in two distinct directions. The plasmon decay products propagating along the same direction may recombine into a plasmon. Therefore in long devices
perturbative treatment of plasmon decay, strictly speaking, is not
justified.

\begin{figure}[ptb]
\includegraphics[width=8.0cm]{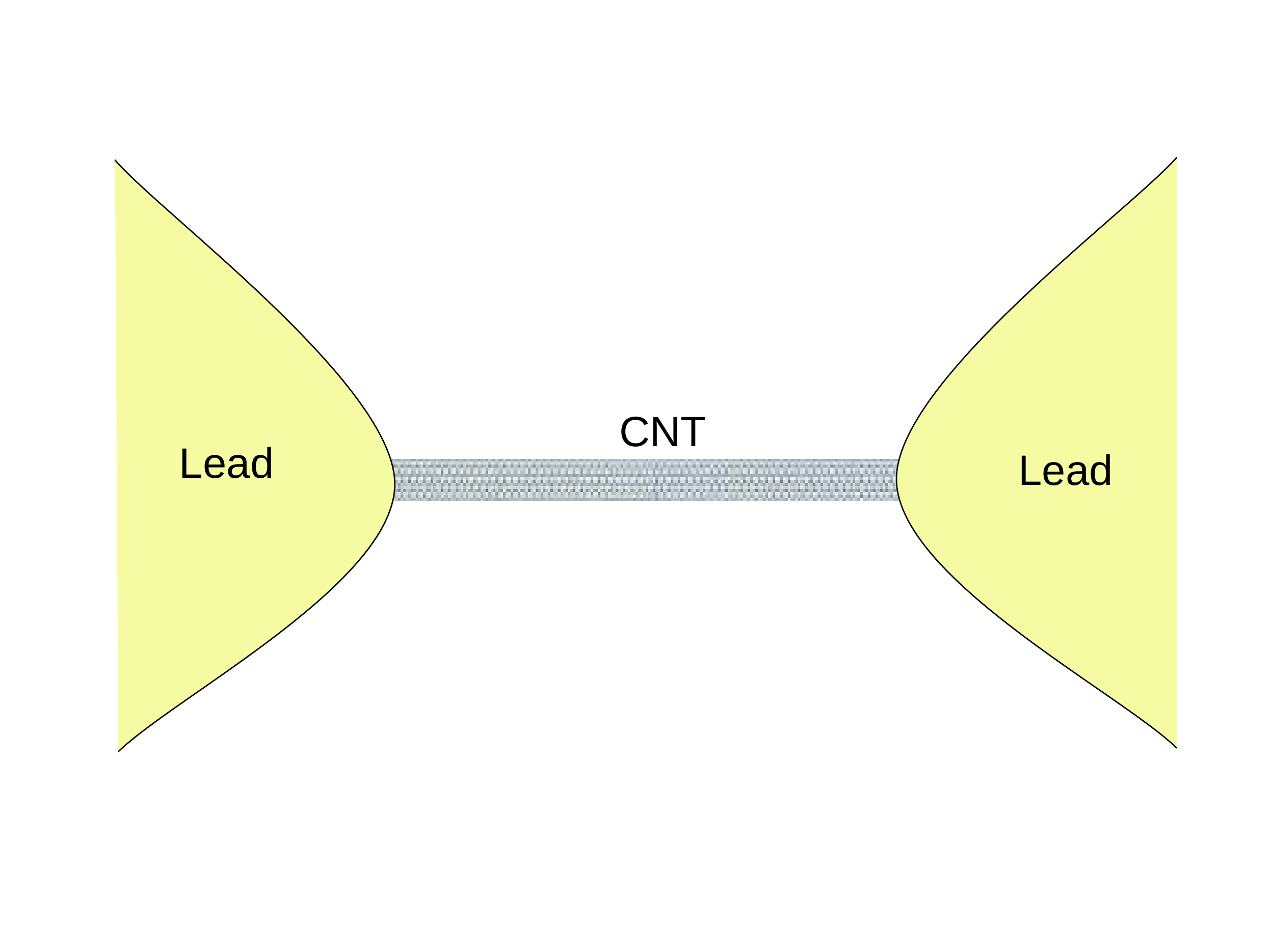}
\caption{(Color online) Schematic representation of the device. A carbon nanotube (CNT) connected to two metal leads is subjected to ac-electric field. } \label{fig:setup}
\end{figure}

In this paper we consider plasmon decay in a finite geometry
illustrated in Fig.~\ref{fig:setup}. In this experimentally
motivated~\cite{Prober} setup a carbon nanotube connected to two
metal leads is subjected to THz electric field. The finite length
of the device enables us to account for the plasmon decay processes
using perturbation theory in the backscattering interactions. We
investigate how the decay processes affect the plasmon resonances and
modify the device conductance.

The paper is organized as follows. In Sec.~\ref{sec:Luttinger} we
present the Luttinger liquid description of electron liquid in
armchair carbon nanotubes. In Sec.~\ref{sec:dissipationless} we obtain
the ac conductance of the device in the absence of plasmon decay
into the neutral modes. In Sec.~\ref{sec:decay} we discuss the physical
mechanisms of plasmon decay and study their influence on energy
dissipation in the ac conductance of the device. Our results are summarized in Sec.~\ref{sec:summary}.

\section{Luttinger liquid description of armchair CNTs \label{sec:Luttinger}}

\begin{figure}[ptb]
\includegraphics[width=9.0cm]{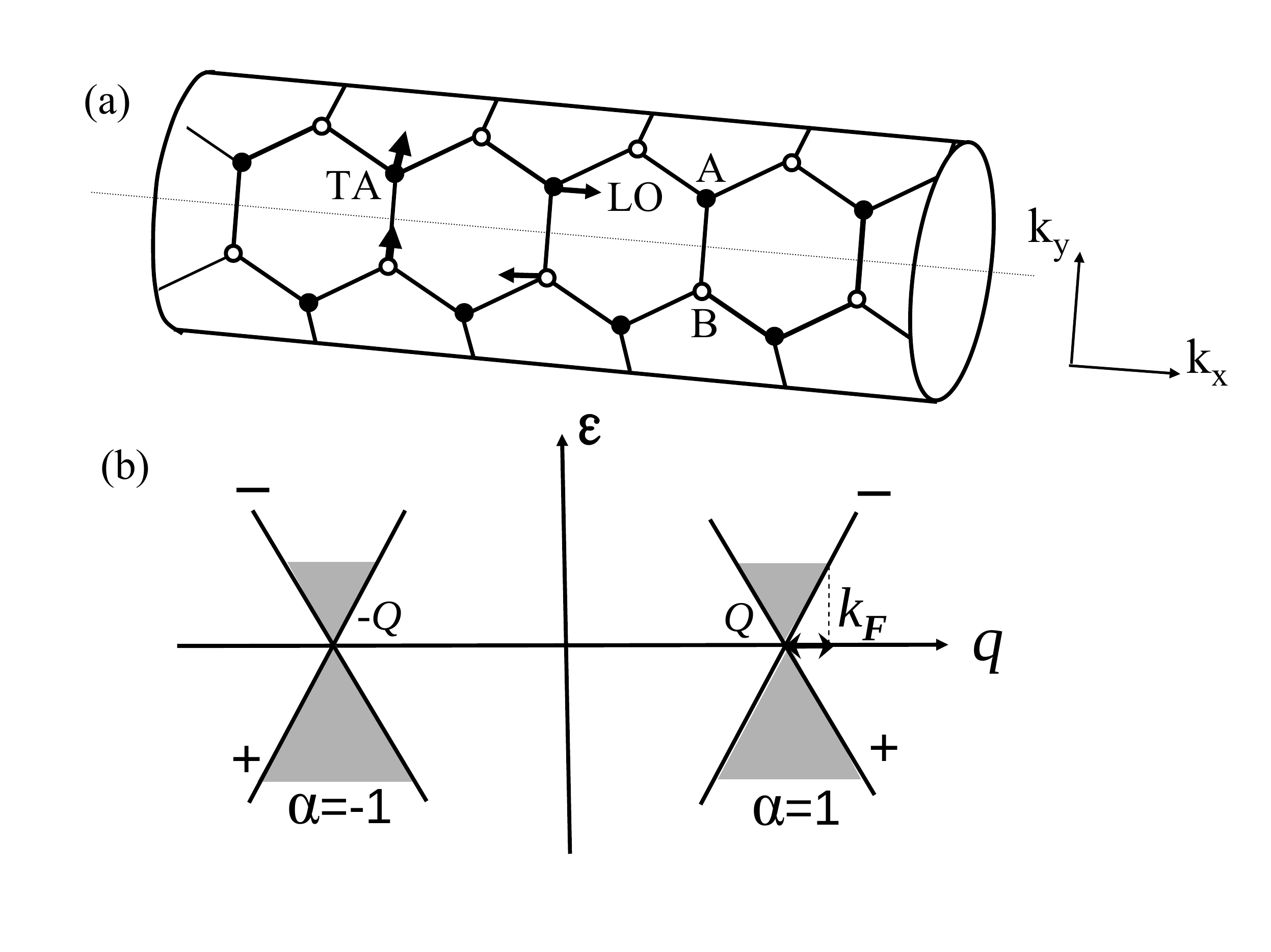}
\caption{(a) Schematic picture of an armchair tube. Full and open
  circles denote atoms in the A and B sublattices. The arrows show
  atomic displacements in the TA and LO phonon modes. (b) Free
  electron spectrum of armchair carbon nanotubes: $+$ and $-$ denote
  the parity of the bands. The two
  valleys are labeled by the index $\alpha= \pm 1$. }. \label{fig:spectrum}
\end{figure}

Near the Fermi level the electron band structure in armchair nanotubes
consists of two spin degenerate bands crossing at two points, $k=\pm
Q$, in the Brillouin zone. Near the crossing points the spectrum has
linear Dirac-like dispersion, see Fig.~\ref{fig:spectrum}. For undoped
tubes the the Fermi level goes through the crossing points. Setting
$\hbar=1$ we write the non-interacting electron Hamiltonian near the Fermi level as
\begin{equation}
H_{0}=-iv_{F}\sum_{\alpha r \sigma}\int dx\, \psi_{\alpha r
\sigma}^{+}(x) \, r\,\partial_{x}\psi_{\alpha r \sigma}(x).
\end{equation}
Here $v_F \approx 8 \times 10^5\,$m$\,$s$^{-1}$ is the Fermi
velocity, $\alpha=\pm 1$ is the valley index, $r=\pm1$ represents
right- and left-  moving electrons, and $\sigma$ is the electron spin.
The electron creation/annihilation operators are denoted by $\psi^\dagger$/$\psi$.
It is important to note that armchair CNT are symmetric about a plane containing
the tube axis and bisecting the AB bonds that point along
the circumference. As a result the electron wave functions can be
classified by parity $P=- \alpha r = \pm 1$, as indicated in
Fig.~\ref{fig:spectrum}. The wave functions with positive/negative
parity are respectively even/odd in the AB sublattices.

We consider an $(N,N)$ armchair CNT, which has $N$ unit cells around
the circumference of the tube, and use $1/N$ as a small parameter in
the theory. For most synthetic CNTs $N\geq 8$, and this is a good approximation.

For $1/N \ll 1$ the forward scattering part of the e-e
interaction is much stronger than the backscattering part, the latter
being small by $1/N$~\cite{Balents1997}. The former can be written as
\begin{equation}
H_{\rho}=\frac{g_{\rho}}{2}\int dx\,  n^2 (x), \quad n(x)=\sum_{\alpha r
\sigma} \psi^{+}_{\alpha r \sigma}(x) \psi_{\alpha r \sigma}(x).
\end{equation}
We assume that the Coulomb interaction is screened, e.g. by a
gate at a distance $d$, which is assumed to be short compared to the length of
the tube, $L$ but longer than the CNT radius $R$. The latter can be expressed in
terms of the graphene lattice constant $a$ as $R=\sqrt{3} N a /2 \pi$. Elementary
electrostatics leads to the estimate
$g_{\rho} \approx2 e^2 \ln (d/R)$.

Thus to zeroth order in $1/N$ backscattering is absent, and
electrons in an armchair tube form a
Luttinger liquid~\cite{Kane1997,Egger1997}. The Hamiltonian is
bosonized by the standard procedure (see, for example,
Refs.~\onlinecite{Giamarchi,TsvelikBook})
\begin{equation}\label{eq:bosonization}
\psi_{\alpha r \sigma}(x)=\frac{F_{\alpha \sigma}}{\sqrt{2\pi \xi}}
\exp\{i[\Theta_{\alpha \sigma}(x)-r\, \Phi_{\alpha
\sigma}(x)+r k_F x]\}.
\end{equation}
Here $F_{\alpha \sigma}$ are the Klein factors, $\xi$ is a short
distance cutoff on the order of the radius of the nanotube, $k_F$ is
the Fermi wavevector counted from the Dirac point and $\Phi$ and
$\Theta$ are bosonic fields satisfying the commutation relation
$[\Phi_{\alpha \sigma}(x),\Theta_{\alpha^{\prime}
\sigma^{\prime}}(x^{\prime})]=-i \delta_{\alpha \alpha^{\prime}}
\delta_{\sigma \sigma^{\prime}} \theta(x-x^{\prime})$,
where $\theta(x-x^{\prime})$ is the step function.

Introducing the symmetric and antisymmetric combinations of charge and
spin modes in the different valleys,
\begin{eqnarray}
\Phi_{\alpha \sigma}&=& \left[ \Phi_{c +} + \alpha \Phi_{c -} + \sigma
\Phi_{s +} +\alpha \sigma \Phi_{s -} \right]/2, \\
\Theta_{\alpha \sigma }&=& \left[ \Theta_{c +} + \alpha \Theta_{c -} + \sigma
\Theta_{s +} +\alpha \sigma \Theta_{s -} \right]/2, \nonumber
\end{eqnarray}
we rewrite the Luttinger liquid Hamiltonian of the CNT including forward
scattering as
\begin{equation}\label{eq:H_forward}
H_{0}+ H_{\rho}= \sum_{j}\frac{ u_{j}}{2\pi}\int dx
{[K^{-1}_{j}(\partial_x \Phi_{j})^2+K_{j}(\partial_x \Theta_{j})^2]}.
\end{equation}
Here $j=c\pm, s\pm$, and the velocity and Luttinger parameter of the
$j$-th mode are related via $u_{j}=v_{F}/K_{j}$. All four bosonic modes
have linear spectrum, $\varepsilon_j=u_j |k|$. For the three neutral modes $j=c-,
s\pm$, the Luttinger parameter $K_j$ is equal to unity and their
velocities coincide with the Fermi velocity. For the charge mode, $j=c_+$, the
Luttinger parameter is significantly smaller than one,
$K_{c+}=1/\sqrt{1+4 g_{\rho}/\pi v_{F}} \approx 0.2$. For the reason,
the plasmon mode has a much higher velocity than that for the three neutral modes. Below we denote the plasmon velocity by $u\equiv v_F/K_{c+}$.

In the absence of decay processes the energy of a plasmon excited in the
device dissipates only due to its emission into the leads. The
power dissipation due to this mechanism is considered in the next
section.

\section{Plasmon resonance in the absence of intrinsic dissipation \label{sec:dissipationless}}

We consider the setup shown in Fig.~\ref{fig:setup}. The plasmons
excited in the tube get partially reflected from the junctions with
the leads. The plasmon reflection processes can be described by a one dimensional
model, Eq.~(\ref{eq:H_forward}) with spatially dependent Luttinger
parameters~\cite{Maslov1995,Safi1995,Ponomarenko1995} defined as
follows. For the charge mode the Luttinger parameter is equal to
$K_{c+}$ inside the tube, $0<x <L$, and unity in the leads. To zeroth
order in $1/N$ the Luttinger parameters in the neutral modes are equal
to unity both in the tube and in the leads.

We denote the THz electric field by $E_\omega (x)$. Its distribution
along the wire depends on the details of the device geometry and is a
solution of the proper electrodynamics problem which is beyond the
scope of this paper; here we assume $E_\omega (x)$ is known. The
dissipation rate of the electric field energy in the device can be
written as
\begin{equation}\label{eq:power}
W(\omega)=\frac{1}{2}\mathrm{Re}\int_0^L dx_1\int_0^Ldx_2 E_{\omega}(x_1)E^*_{\omega}(x_2)\sigma_{\omega}(x_1,x_2),
\end{equation}
where $\sigma_\omega(x,x')$ is the nonlocal conductivity, which can be
expressed via the retarded current-current correlation function by the
usual Kubo formula,
\begin{equation}\label{eq:Sigma}
\sigma_{\omega}(x,x')=\frac{e^2}{\omega}\int_{0}^{\infty}dt\, e^{i\omega t}\langle[j(x,t),j(x',0)]\rangle .
\end{equation}

We assume that the electric field (\emph{i}) varies smoothly on the
scale of the lattice constant within the tube and (\emph{ii}) preserves
the reflection symmetry of the tube. Because of (\emph{i}) it will not
scatter electrons between the valleys, which involves momentum
transfer $\sim 2Q$. Due to \emph{ii}) parity conservation also forbids
backscattering of electrons within the same valley. Thus the electric
field couples only to the smooth part of the current density, which
can be expressed in the bosonized form as $j=2\frac{e}{\pi}\partial_t
\Phi_{c+}$. Therefore we can express the conductivity in terms of the
Green's function of the $\Phi_{c+}$ field,
\begin{equation}\label{eq:conductivity_def}
\sigma_{\omega}(x,x')=4e^2\frac{i\omega}{\pi^2}G^R_{\omega}(x,x'),
\end{equation}
where
\begin{equation}\label{eq:G_{omega}}
G^R_{\omega}(x,x')=-i\int_{0}^{\infty}dt\, e^{i\omega t}\langle [\Phi_{c+}(x,t),\Phi_{c+}(x',0)]\rangle
\end{equation}
is the retarded GF of the plasmon mode.

We denote the Green's function of the plasmon mode in the absence of
backscattering by $G^R_{0,\omega}(x,x')$. It satisfies the
differential equation
\begin{equation}\label{eq:GF}
\left\{\partial_x \left(\frac{v_F}{K^2(x)}\partial_x\right)+\frac{\omega^2}{v_F}\right\}G^R_{0,\omega}(x,x')=\pi \delta(x-x'),
\end{equation}
with the boundary conditions: (1) $G^R_{0,\omega}(x,x')$ is continuous
at $x=0,L$ and $x=x'$; (2) $[v_F/K^2(x)]\partial_x
G^R_{0,\omega}(x,x')$ is continuous at $x=0,L$.  For our model
($K(x)$ equal to $K_{c+}$ inside the tube and unity outside) the
Green's function can be expressed in terms of the reflection
amplitude, $\mathrm{r}=\frac{1-K_{c+}}{1+K_{c+}}$, of the plasmon from the
junction with the lead.

The dissipated power in Eq.~(\ref{eq:power}) depends only on the
Green's function inside the tube. In this region for $0<x<x'<L$ it is
given by
\begin{eqnarray}\label{eq:Solution}
G^R_{0,\omega}(x,x')&=&\frac{\pi K_{c+}}{2i\omega}
\frac{e^{i\omega L/u}}{ 1-\mathrm{r}^2 e^{2i\omega L/u}}
\left( e^{-i\frac{\omega}{u}x}+\mathrm{r}\,  e^{i\frac{\omega}{u}x}\right) \nonumber \\
& & \times \left( e^{i\frac{\omega}{u}(x'-L)}+\mathrm{r}\,  e^{-i\frac{\omega}{u}(x'-L)}\right).
\end{eqnarray}
In the region $0<x'< x< L$ the Green's function is obtained by
exchanging $x$ and $x'$ in the right hand side.

The Green's function above has poles as a function of frequency
$\omega$. The real part of the pole location gives the resonance
frequency $\omega_n=n\pi u/L$, and its imaginary part gives the
resonance half-width, $\gamma_Z$.  The plasmon reflection amplitude
from the junctions with the leads is determined by the ratio of the
plasmon velocities in the tube and in the lead,
$\mathrm{r}=\frac{1-K_{c+}}{1+K_{c+}}$.  This gives the escape half-rate
$\gamma_Z=\frac{u}{L}\log{\left(\frac{1+K_{c+}}{1-K_{c+}}\right)}$.
In the limit of high charge stiffness, $K_{c+}\ll 1$, it becomes  much
smaller than the spacing between consecutive resonances and independent of the interaction strength, $\gamma_z\simeq
2K_{c+}\frac{u}{L}=2v_F/L$.

The dissipated power is given by Eqs.~(\ref{eq:power}),
(\ref{eq:conductivity_def}), and (\ref{eq:Solution}) and depends not
only on the properties of the tube but also on the spatial
distribution of the electric field (as mentioned above, the latter
depends on the geometry and is assumed known).

\section{Plasmon decay in the device \label{sec:decay}}

The energy dissipation described by the ac conductance obtained in
Sec.~\ref{sec:dissipationless} corresponds to the emission of plasmon
waves into the leads. The THz electric field excites plasmon modes of the
electron fluid, which after a few reflections from the contacts with
the leads are radiated out into the leads. Thus the energy absorbed
from the THz radiation is re-radiated into the leads in the form of a
plasmon flux.

In the presence of backscattering interactions, there is an additional
contribution to the dissipation energy in the device. The
plasmon waves generated by the THz radiation can decay into neutral
modes before escaping into the leads. In this dissipation channel the
absorbed energy is carried into the leads by the neutral modes. We
discuss the microscopic interactions responsible for the decay of
plasmons into the neutral modes in
Sec.~\ref{sec:backscattering_interactions} and study how these
processes modify the device conductance in
Sec.~\ref{sec:intrinsic_dissipation}.

\subsection{backscattering processes leading to plasmon decay \label{sec:backscattering_interactions}}

The possible backscattering processes obey certain symmetry selection
rules. In particular parity conservation forbids single electron
backscattering in armchair carbon nanotube due to e-e Coulomb interaction, which also attributes to the Klein paradox in graphene~\cite{Geim} but two, or more generally, any even number
of particles can be back-scattered. The electron interaction
Hamiltonian contains such processes but their dimensionless coupling
constants are small in $1/N$ and can be treated perturbatively. Having
in mind temperatures well below the gap between the one-dimensional
subbands ($v_F/R \sim 1eV$) we retain only the most relevant
backscattering interactions. Namely the processes, which scatter two right
movers into two left movers or vice-versa~\cite{Kane1997,Egger1997,Odintsov1999}. Although for the Brillouin zone shown in Fig.~\ref{fig:spectrum} b) such scattering conserves quasimomentum and thus corresponds to normal processes, we refer to it as Umklapp scattering, following the accepted nanotube terminology~\cite{Kane1997,Egger1997,Odintsov1999}.

In the bosonized representation the Umklapp Hamiltonian density can be
written as~\cite{Nersesyan2003}
\begin{eqnarray}\label{eq:H_U}
 \mathcal{H}_{u}&=&- \, \frac{\cos(2\Phi_{c+}\!-\!4k_F x)}{2(\pi \xi)^2}
               \{g_3 \cos(2\Theta_{s-})
                +\nonumber \\
                 &&(g_3-g_1)\cos(2\Phi_{s+})+ \nonumber \\
                 && g_1[\cos(2\Phi_{c-})-
                \cos(2\Phi_{s-})]\}.
\end{eqnarray}
where the coupling constants originate from the short range part of
the Coulomb interaction. They can be expressed in terms of the matrix
elements of Coulomb interaction between electrons residing in the $A$
and $B$ sublattices, $V_{AA}(q)$ and $V_{AB}(q)$, as follows:
$g_3=V_{AA}(2Q)+V_{AB}(2Q)$ and
$g_1=V_{AA}(0)-V_{AB}(0)$~\cite{Odintsov1999,Nersesyan2003}. Here $2Q$
is the momentum difference between the two Dirac points, see
Fig.~\ref{fig:spectrum}, and we assumed that the Fermi momentum due to
finite doping is small, $k_F\ll Q$.  Both $g_1$ and $g_3$ arise from
distances on the order of the lattice constant and are on the order of
$e^2/N$.

In the presence of phonons single electron backscattering can occur because
of electron-phonon interaction. The phonon modes in carbon
nanotubes are classified to two types by the parity symmetry: \emph{i)}
transverse acoustic (TA), longitudinal optical (LO), and radial optical
(RO) with negative parity, and \emph{ii)} transverse optical (TO), longitudinal
acoustic (LA), and radial acoustic (RA) with positive
parity~\cite{Jishi,Mahan,Chen}. The couplings of radial phonon modes to
electrons are weaker in $1/N$~\cite{Mahan,Chen}
than those for other phonon modes. Therefore  we neglect their effects here. For the
remaining couplings, one can distinguish between inter-valley and
intra-valley scattering. The former requires a phonon with a momentum
on the order of the inverse lattice spacing. Such phonons have high
frequencies $\sim 10^3 K$ and are frozen out at room temperature. We
thus neglect such processes and consider only intra-valley
backscattering.  Such processes change the parity of the electron
wavefunction and thus require phonons with negative parity, i.e., TA
or LO phonons. Their displacements are illustrated in
Fig.~\ref{fig:spectrum}. Since the frequency of the LO phonon is $\sim
1000 K$ it is frozen out in the temperature range we are interested
in. Thus below we restrict our treatment to the backscattering of electrons from the
TA phonons only.

The Hamiltonian of the TA phonon mode is given by
\begin{equation}
H_{TA}=\frac{1}{2}\int dx \left[\frac{1}{s_T}(\dot{\varphi}_T(x))^2+s_T(\partial_x \varphi_T(x))^2\right].
\end{equation}
Here $\varphi_T(x)=\sqrt{\rho s_T}u_T(x)$ is the dimensionless TA
phonon field, $\rho$ the linear mass density along the tube axis,
$u_T(x)$  - the atomic displacement of the TA phonon mode, and $s_T \sim
1.4\times 10^4$ m$/$s is the phonon velocity.
The deformation of the TA mode couples to the time-reversal invariant
operator bilinear in the fermions $M(x)=-i \sum_{\alpha r
  \sigma}\alpha r \psi^{\dagger}_{\alpha r \sigma}(x) \psi_{\alpha -r
  \sigma}$,
\begin{equation}\label{eq:ep_fermion}
H_{ep}=  \int dx M(x)   g_T \partial_x \varphi_{T}(x).
\end{equation}

In the bosonized representation the Hamiltonian density for the electron-TA
phonon coupling can be written as~\cite{Chen}
\begin{eqnarray}\label{eq:H_{eph}}
\mathcal{H}_{ep}&=& - \frac{4 g_T}{\pi
 \xi} \partial_x \varphi_{T}\!\left[\prod_{\nu=\pm}\cos(\Phi_{c+}\!-\!2k_F x)\cos{(\Phi_{c-})}
 \sin(\Phi_{s\nu})  \right. \nonumber \\
 && + \left.  \prod_{\nu=\pm}\sin(\Phi_{c+}-2k_F x)\sin{(\Phi_{c-})}
 \cos(\Phi_{s\nu}) \right] ,
\end{eqnarray}
where $g_T$ is the e-ph coupling constant. Within the tight-binding
model, the coupling constant $g_T$ can be expressed in terms of the
derivative of the transfer integral $J(r)$ with respect to the bond
length $r$, see Ref.~\onlinecite{Jishi}. In an armchair carbon
nanotube, $g_T=\frac{\sqrt{3}}{4} \frac{a}{\sqrt{\rho s_T}}\left.
  \frac{\partial J(r)}{\partial r}\right|_0$ and for $(N,N)$ armchair
nanotubes, $\rho=4NM/a$, where $M$ is carbon atom mass and $a$ is the
lattice constant of a graphene sheet.

\subsection{intrinsic energy dissipation in the device and plasmon decay \label{sec:intrinsic_dissipation}}

The backscattering interactions lead to the possibility of plasmon
decay into the neutral modes. Using the fact that the backscattering
interactions are small in $1/N$ we evaluate the decay rate and the
corresponding correction to the device conductance using perturbation
theory.  We work in the regime where the cumulative plasmon decay rate
in the device is smaller than the rate of elastic escape of plasmons
into the leads. In this case the correction to the device conductance
due to plasmon decay can be treated perturbatively.

Both the Umklapp and the electron-phonon backscattering interactions
are expressed in terms of exponentials of the charge mode,
$\Phi_{c+}$. As a result, their contribution to the plasmon decay and
the device conductance can be evaluated along similar lines.

To second order in perturbation theory there is no cross-correlation
between the Umklapp and electron phonon backscattering processes, and
the retarded Green's function of the charge mode can be expressed as
\begin{eqnarray}\label{eq:G^R}
G^R_{\omega}(x,x')&=&G^R_{0,\omega}(x,x')+\nonumber \\
&&\sum_\eta\int^{L}_0 dx_1\int^{L}_0 dx_2\,
G^R_{0,\omega}(x,x_1)  \times \nonumber \\
&&\mathcal{K}^{\omega}_\eta(x_1,x_2)G^R_{0,\omega}(x_2,x'),
\end{eqnarray}
where the index $\eta$ denotes either the Umklapp ($\eta=u$) or the
electron-phonon ($\eta=ep$) interaction, and the kernel
$\mathcal{K}^{\omega}_\eta(x_1,x_2)$ can be expressed in terms of the
retarded correlation function of the corresponding backscattering
perturbation as
\begin{eqnarray}\label{eq:kernel}
\mathcal{K}^{\omega}_\eta(x,x')&=& -i a_\eta^2 \int dt \left[ e^{i\omega t}\langle\left[\mathcal{H}_\eta(x,t),\mathcal{H}_\eta(x',0)\right]\rangle_0 - \right. \nonumber \\
&&\left. \!\int_0^L \!d\xi\langle \left[\mathcal{H}_\eta(x,t),\mathcal{H}_\eta(\xi,0)\right]\rangle_0 \delta(x-x')\right].
\end{eqnarray}
Here $a_\eta$ is the coefficient in the exponential dependence, $\sim
(e^{ia_\eta \Phi_{c+}} + h.c.)$, of the corresponding backscattering
perturbation on the charge field $\Phi_{c+}$. The derivation of Eqs.~(\ref{eq:G^R}) and (\ref{eq:kernel}) is outlined in Appendix~\ref{sec:fusion}.  The
first term in Eq.~(\ref{eq:G^R}) is the Green's function of the
plasmon in the Luttinger liquid approximation,
Eq.~(\ref{eq:Solution}). The remaining terms represent correction due
to backscattering interactions. The first term in the right hand side of
Eq.~(\ref{eq:kernel}) corresponds to the usual RPA-like diagram, the
second term comes from the tadpole diagram and describes elastic
scattering. The average $\langle \ldots \rangle_0$ is performed with
respect to the quadratic Luttinger liquid Hamiltonian,
Eq.~(\ref{eq:H_forward}).

In the spatially inhomogeneous setup considered here the kernel $\mathcal{K}^\omega_\eta(x_1,x_2)$ depends not only on the coordinate
difference but also on the center of mass coordinate $(x_1+x_2)/2$. Its evaluation leads to rather cumbersome expressions. The situation simplifies in the physically relevant regime,
where the characteristic thermal wavelength of the neutral modes, $L_T =v_F/T$, is
much shorter than the size of the device $L$, while the wavelength of the plasmon resonance is on the order $L$.

In this case the Green's functions of the
plasmon mode in the second term of Eq.~(\ref{eq:G^R}) depend on $x_1$
and $x_2$ on the scale of wavelength of the plasmon resonance, $\sim
L$. On the other hand, because backscattering operators depend
exponentially on the neutral modes, $H_\eta(x,t)\sim \exp (i a_\eta
\Phi_j(x,t))$, their correlator $\mathcal{K}^\omega_\eta(x_1,x_2)$
falls off exponentially for $|x_1-x_2| > L_T$, which is much shorter
than the length of the tube L.  This is especially apparent in the
case of a uniform wire for which the correlators of exponentials of
bosonic fields have a simple form~\cite{Giamarchi},
\begin{widetext}
\begin{equation}\label{eq:exponential_correlator}
   \left\langle e^{i a \Phi_j(x,t)} e^{-i a  \Phi_j(0,0)}\right\rangle_0 \equiv
     \Pi(K_j, a; x,t)= \frac{\left( \pi \xi T K_j/v_F\right)^{a^2  K_j/2}}{\left\{ \sinh[\pi T(K_j x/v_F-t_-)]\sinh[\pi T(K_j x/v_F+t_-)]\right\}^{a^2  K_j /4}},
\end{equation}
\end{widetext}
where $t_-=t-i\epsilon$.

In the inhomogeneous situation the kernel $\mathcal{K}^\omega_\eta(x_1,x_2)$ still decays exponentially for $|x_1-x_2| > L_T$. Therefore the main contribution to the Green's function correction in Eq.~(\ref{eq:G^R}) comes from the region $|x_1-x_2| \lesssim L_T$.  At such
separations only the fluctuations of bosonic modes with wavelengths less than
$L_T$ contribute to the correlator. In the regime of interest, $L_T\ll
L$, such short wavelength vibrations can be described semiclassically
and are insensitive to the boundary conditions at the junctions, which
describe reflection processes~\footnote{The quasiclassical treatment
  becomes inapplicable in the regions within $L_T$ of one of the
  junctions, but the contribution of these regions to the overall
  decay rate is small as compared to that of the rest of the tube.}.
Thus we come to the conclusion that in the bulk of the nanotube, i.e.
with the exception of the regions of size $L_T$ near the junctions the
correlator $\mathcal{K}^\omega_\eta(x_1,x_2)$ is the same as in a
homogeneous tube, and can be replaced by the local kernel
\begin{equation}\label{eq:K_locality}
    \mathcal{K}^\omega_\eta(x_1,x_2) \to \delta(x_1-x_2)\mathcal{K}^{\omega}_\eta(q=0),
\end{equation}
where
\begin{eqnarray}\label{eq:K}
\mathcal{K}^{\omega}_\eta(q=0)&=& -i a_\eta^2\int_{-\infty}^\infty dx \int_0^\infty dt (e^{i\omega t}-1) \nonumber\\
&& \times \langle\left[\mathcal{H}_\eta(x,t),\mathcal{H}_\eta(0,0)\right]\rangle_0.
\end{eqnarray}

This observation enables us to express the correction to the
dissipated power due to decay of plasmons into neutral modes in terms
of intrinsic parameters of uniform nanotubes. Accordingly, below all
correlation functions of backscattering operators will be evaluated
for infinite homogeneous nanotubes.

With the aid of Eq.~(\ref{eq:K_locality})
the power dissipation rate in Eq.~(\ref{eq:power}) can be written in the form
\begin{equation}\label{eq:powerreducedform}
W(\omega)=W_0(\omega)+\rm{Im}\sum_\eta \mathcal{K}_\eta^\omega(q=0) F(\omega).
\end{equation}
Here the first term corresponds to the dissipation rate due to the escape
of plasmon to the leads and is given by
\begin{eqnarray}
W_0(\omega)&=& \frac{2 e^2\omega}{\pi^2}\int_0^L dx_1\int_0^L dx_2 \nonumber\\
&& \times \,{\rm{Im}}\, E_{\omega}(x_1)E^*_{\omega}(x_2)G^R_{0,\omega}(x_1,x_2).
\end{eqnarray}
The second term in Eq.~(\ref{eq:powerreducedform}) corresponds to the intrinsic dissipation due to the electronic Umklapp processes and electron-phonon backscattering. It can be expressed in terms of kernel for an infinite tube, Eq.~(\ref{eq:K}), and the  form-factor $F(\omega)$,
\begin{eqnarray}\label{eq:formfactor}
F(\omega)&=&\frac{2e^2\omega}{\pi^2}\int_0^L dx_1 \int_0^L dx_2 \int_0^L dx_3 \, E_{\omega}(x_1) \, E^*_{\omega}(x_3) \nonumber\\
&& \times\,  G^R_{0,\omega}(x_1,x_2)\, G^R_{0,\omega}(x_2,x_3),
\end{eqnarray}
which is quadratic in the THz electric field and depends, through its spatial distribution, on the geometry of the device.

Both the electric field and the unperturbed Green's function $G^R_{0,\omega}$ of the Luttinger liquid are temperature independent. Therefore $W_0(\omega)$ and the form-factor $F(\omega)$ do not depend on temperature. All the temperature dependence of the dissipation rate
$W(\omega)$ is thus totally determined by intrinsic properties of a uniform CNT and is described by the kernel $\mathcal{K}_\eta^\omega(q=0)$.

We note that the kernel $\mathcal{K}^{\omega}_\eta(q=0)$ also determines the
ac-conductivity $\sigma_\omega(q=0)$ of an infinite
wire~\cite{Giamarchi} in the high frequency or high temperature regime, where
back-scattering interactions can be treated perturbatively. The latter can be expressed in the form
\begin{equation}\label{eq:conductivity}
\sigma(\omega,T)=\frac{4e^2 K_{c+}u}{\pi (-i\omega + \sum_\eta\gamma_{\eta})}.
\end{equation}
Here $\gamma_{\eta}$ is the relaxation rate due to the Umklapp ($\eta=u$) or electron-phonon ($\eta=ep$) back-scattering, which can be expressed in terms of the local kernel as
\begin{equation}\label{eq:Gamma}
\gamma_{\eta}\equiv\frac{i}{\omega} K_{c+}u_{c+}\, \mathcal{K}^\omega_\eta(q=0).
\end{equation}

Thus with the aid of Eqs.~(\ref{eq:powerreducedform}), (\ref{eq:formfactor}) and (\ref{eq:Gamma}) the  temperature dependent part of the power dissipation in the device can be expressed in terms of the intrinsic plasmon decay rate in a uniform CNT, $\gamma_\eta$, which is given by Eqs.~(\ref{eq:Gamma}) and (\ref{eq:K}).

In the remainder of this section we evaluate the intrinsic plasmon decay rate due to various backscattering processes.

\subsubsection{Umklapp mechanism of attenuation \label{sec:Umklapp}}

We begin by considering Umklapp-mediated plasmon decay rate, $\gamma_u$.
The Umklapp Hamiltonian density in Eq.~(\ref{eq:H_U}) contains
exponential operators of both $\Phi$- and $\Theta$-fields
corresponding to the neutral modes. Because the Luttinger parameter
$K$ for neutral modes is equal to unity, the correlators of
exponential operators of $\Phi$- and $\Theta$-fields coincide. As a
result we obtain
\begin{eqnarray}\label{eq:Correlator}
\langle \mathcal{H}_{u}(x,t)\mathcal{H}_{u}(0,0)\rangle_0&=&\frac{3g^2_1+2g^2_3-2g_1g_3}{16(\pi \xi)^4}\,\cos{(4k_F x)} \times \nonumber \\
&& \Pi(K_{c+},2;x,t)\Pi(1,2;x,t).
\end{eqnarray}
The analogous correlator corresponding to the opposite ordering of backscattering operators can be obtained from the identity
\[
\langle \mathcal{H}_{u}(0,0)\mathcal{H}_{u}(x,t)\rangle_0=\langle \mathcal{H}_{u}(x,t)\mathcal{H}_{u}(0,0)\rangle^*_0.
 \]

Using Eqs.~(\ref{eq:Gamma}), (\ref{eq:K}), (\ref{eq:exponential_correlator}), and  (\ref{eq:Correlator}) we obtain
\begin{equation}\label{eq:gamma_u_main}
\gamma_{u}=\left(\frac{\pi T\xi }{ u}\right)^{2K_{c+}}\frac{3g^2_1+2g^2_3-2g_1g_3}{4 T (\pi\xi)^2} \, F_u\left(\frac{\omega}{T},\frac{v_F k_F}{T}\right),
\end{equation}
where $F_u(\frac{\omega}{T},\frac{v_F k_F}{T})$ is a scaling function depending on the ratio $\omega/T$ and $v_F k_F/T$. It is given by the following double integral,
\begin{widetext}
\begin{eqnarray}\label{eq:F}
F_u\left(\frac{\omega}{T},\frac{v_F k_F}{T}\right)&=&\frac{T}{\omega}\int_{-\infty}^{\infty}dX\int_0^{\infty}dY \left(e^{i\frac{\omega}{T}Y}-1\right)
\cos{\left(\frac{4k_Fv_F}{T}X\right)}\nonumber\\
&&\left[\frac{1}{(\sinh{\pi(\frac{v_f}{u}X+Y_-)}
\sinh{\pi(\frac{v_f}{u}X-Y_-)})^{K_{c+}}}\frac{1}{\sinh{\pi(X+Y_-)}
\sinh{\pi(X-Y_-)}}-c.c.\right],
\end{eqnarray}
\end{widetext}
where $Y_-\equiv Y-i\epsilon$.

Using the standard deformation of the integration contour, $Y_-\to
Y-i/2$, in the above equation one may express the real part of the scaling
function in the form that is more convenient for numerical evaluation,
\begin{widetext}
\begin{equation}\label{eq:F_u_real}
\Re F_u\left(\frac{\omega}{T},\frac{v_F k_F}{T}\right)=\frac{T}{\pi^2 \omega}\sinh \frac{\omega}{2 T} \int_{-\infty}^{\infty}d x\int_{-\infty}^{\infty} d y \frac{\cos{\left(y \, \omega/\pi T \right)} \cos{\left(4 k_F v_F x/\pi T \right)}}{\left[\cosh{(K_{c+} x+y)}
\cosh{(K_{c+}x-y)}\right]^{K_{c+}} \cosh{(x+y)}\cosh{(x-y)}}.
\end{equation}
\end{widetext}
In the limit of high charge stiffness, $K_{c+}\to 0$, the above expression simplifies to,
\begin{equation}\label{eq:F_u_Kc0}
    \Re F_u\left(\frac{\omega}{T},\frac{v_F k_F}{T}\right)=  \frac{\frac{T}{2\omega} \sinh{\frac{\omega}{2T}}}{\cosh\left(\frac{4k_F v_F+\omega}{4T}\right)\cosh\left(\frac{4k_F v_F-\omega}{4T}\right)}.
\end{equation}

\begin{figure}[ptb]
\includegraphics[width=8.0cm]{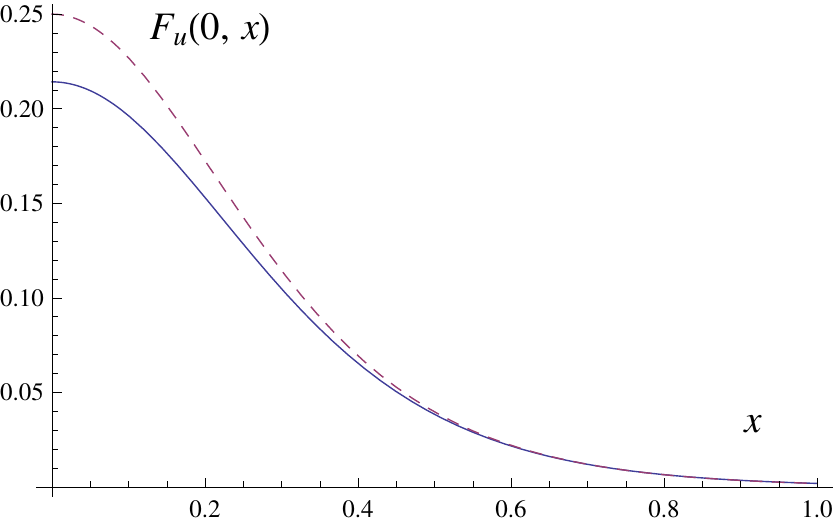}
\caption{Dependence of the scaling function in the static limit, $F_u(0, x)$ on the doping parameter $x=\frac{k_F v_F}{T}$ for $K_{c+}=0.2$ (solid line) and $K_{c+}=0$ (dashed line).} \label{fig:F_u_plot}
\end{figure}

In the temperature regime of interest, $T\gg \omega \sim u/L$,
the function $F_u(\frac{\omega}{T},\frac{v_F k_F}{T})$ in
Eq.~(\ref{eq:F}) reaches the dc limit $F_u(0,\frac{v_F k_F}{T})$,
which is purely real and depends only on the doping level.  For
comparison, in Fig.~\ref{fig:F_u_plot} we plot the static limit of the
scaling function $F_u(0,k_F v_F/T)$ evaluated numerically at
$K_{c+}=0.2$ using Eq.~(\ref{eq:F_u_real}) and the analytical
expression from Eq.~(\ref{eq:F_u_Kc0}) for $K_{c+}=0$.

At small doping, the scaling function goes to a constant and
$\gamma_u$ exhibits a power law dependence on the temperature,
$\gamma_u \sim T^{2K_{c+}-1}$. At strong doping, $2 k_F v_F > T$ the
scaling function acquires an activated temperature dependence. This
results in exponential decay of $\gamma_u \sim T^{2K_{c+}-1}
\exp(-2k_F v_F/T)$. The physical reason for the exponential
temperature decay can be understood from simple inspection of energy and
momentum conservation, which shows that energetically optimal process involves backscattering of two thermal quasiparticles exactly at the Dirac
point $k=Q$. At strong doping, $k_F v_F \gg T$
the occupancy of such states is exponentially small, $\sim \exp(-2 k_F
v_F/T)$.

Equation (\ref{eq:gamma_u_main}) and the subsequent equations express
the plasmon decay rate $\gamma_u$ in terms of the somewhat arbitrary
short distance cutoff, $\xi$, which has the order of the radius of carbon nanotube. Alternatively, $\gamma_u$ can be
expressed in a renormalized form as a function of an intrinsic
cutoff-independent low energy scale. In the case of Umklapp
interactions the role of such a low energy scale is played by the
Umklapp gap. Namely, the Umklapp interaction is a relevant
perturbation relative to the Luttinger liquid fixed point and drives
the system to a Mott insulating state with the
gap~\cite{Odintsov1999},
\[
\Delta_u \sim \frac{v_F}{\xi} \left( \frac{g_u}{v_F} \right)^{1/(1-K_{c+})}.
\]
Here $g_u \sim e^2/N$ is the strength of the Umklapp interaction. This gap depends on the radius of the CNT as $R^{-[1+1/(1-K_{c+})]}$ and is hard to evaluate theoretically due to the ambiguity of the coupling constant, but numerical results estimate that it's of the order of $1 - 10$ meV for an armchair CNT of radius $\sim 1$ nm~\cite{Odintsov1999}.  For the perturbation calculation to be applicable, the temperature we consider should be above the Umklapp gap. Up to a numerical coefficient on the order unity the plasmon decay rate
in Eq.~(\ref{eq:gamma_u_main}) can be expressed in terms of this gap
as \begin{equation}\label{eq:gamma_u_renormalized} \gamma_u\sim
  \Delta_u \left( \frac{T}{\Delta_u} \right)^{2K_{c+}-1}
  F_u\left(\frac{\omega}{T},\frac{v_F k_F}{T}\right).
\end{equation}
At zero doping, $\gamma_u \sim T$ for $K_{c+}\to 1$, i.e., without considering the Luttinger liquid effect. Yet in the Luttinger liquid description with strong Coulomb interaction, $K_{c+}\to 0$, $\gamma_u\sim \Delta^2_u/T$.

Note that according to Eqs.~(\ref{eq:conductivity}) and (\ref{eq:gamma_u_renormalized}) the Umklapp processes yield $\sigma (\omega=0, T)\propto T^{1-2K_{c+}}$ at zero doping and high temperature, which is consistent with the result obtained in a previous work~\cite{Kane1997}. In the case $K_{c+}\to 1$, i.e., in the absence of Luttinger liquid effects, the decay rate due to Umklapp processes $\gamma_u\sim T$, and the resistivity has linear dependence on temperature at high temperature, which is consistent with the perturbative calculation of plasmon decay rate from Umklapp processes in CNTs~\cite{unpublished, Balents1997}. We can see that the inclusion of the long range interaction significantly modifies the dependence of the plasmon decay rate on temperature due to Umklapp processes. In the case of strong Coulomb interaction, i.e., $K_{c+}\to 0$, the temperature dependence of decay rate is strongly modified and becomes $\gamma_u\sim T^{-1}$ at high temperature.

\subsubsection{Phonon-assisted mechanism of attenuation \label{sec:e-ph}}

We now consider plasmon decay rate due to electron-phonon interactions, $\gamma_{ep}$. Its physical mechanism can be
understood as follows. Parity-breaking deformations in the TA mode
cause single particle backscattering, which causes the plasmon decay.
Due to the presence of the adiabatic parameter $s_T/v_F \sim 2 \times
10^{-2}\ll 1$ the lattice deformations are effectively static. Thus the energy of the decaying plasmon is transferred into the neutral modes of the electron liquid.
Therefore this mechanism,
which we analyze below, may be characterized as phonon-assisted plasmon
decay. In two dimensions it has been studied in Ref.~\onlinecite{Mishchenko2004}.
At finite doping single electron backscattering involves phonons with momentum
$2k_F$. With the exception of very low temperatures, $T < 2k_F s_T$,
the occupation numbers of such phonons are large, $\sim T/2k_Fs_T$.
This corresponds to classical thermal fluctuations of the lattice.
Since the probability of such fluctuations is determined by the elastic
energy $\sim \exp(-s_T\int d x [\partial_x \phi_{T}(x)]^2/2T)$, the
one-electron backscattering potential $\sim
\partial_x \phi_{TA}(x)$ (see Eq.~(\ref{eq:ep_fermion})) may be viewed as
$\delta$-correlated in space. More precisely, the correlator of
backscattering potential $\sim
\partial_x \phi_{TA}(x)$ becomes exponentially small at distances
greater than $s_T/T$, as is clear from Eq.~(\ref{eq:Phonon}) below. As
the electron correlators depend on the spatial coordinates on much larger
scales $\sim L_T \gg s_T/T$ the correlator of deformation potentials becomes effectively short range. In this respect the phonon-assisted plasmon decay mechanism is similar to that induced by a short range disorder potential. The temperature dependence of disorder-induced conductivity was obtained in Ref.~\onlinecite{Yoshioka}, $\sigma_{dis}\sim \langle V^2 \rangle T^{(K_{c+}-1)/2}$, where $\langle V^2 \rangle$ is the variance of the random potential. In contrast to the temperature-independent variance of the disorder potential, the variance of the phonon correlator scales linearly with the temperature. This gives the power law $\gamma_{ep}\sim T^{(1+K_{c+})/2}$, which is confirmed by the calculations below.

The phonon-assisted decay rate $\gamma_{ep}$ can be evaluated
similarly to the case of Umklapp processes. Using
Eq.~(\ref{eq:exponential_correlator}) we obtain the following
expression for the correlation function for the electron-phonon
Hamiltonian densities given by Eq.~(\ref{eq:H_{eph}}),
\begin{eqnarray}\label{eq:Eph}
 \langle \mathcal{H}_{ep}(x,t)\mathcal{H}_{ep}(0,0)\rangle & = & 2 \left(\frac{g_T}{\pi\xi}\right)^2 \cos{(2k_F x)}D_{ph}(x,t) \nonumber \\
&& \times\, \prod_j \Pi(K_j,1;x,t),
\end{eqnarray}
where $j=c\pm, s\pm$, and $D_{ph}(x,t)\equiv \langle \nabla
\varphi_T(x,t)\nabla \varphi_T(0,0)\rangle$ is the phonon correlator
given by
\begin{eqnarray}\label{eq:Phonon}
D_{ph}(x,t)&=&-\sum_{\nu=\pm 1}\frac{(\pi T/ s_T)^2}{2\sinh^2[\pi T(\frac{x}{s_T}+\nu t_-)]}.
\end{eqnarray}
The correlator with opposite ordering can be obtained from
Eq.~(\ref{eq:Eph}) using the identity $\langle
\mathcal{H}_{ep}(0,0)\mathcal{H}_{ep}(x,t)\rangle=\langle
\mathcal{H}_{ep}(x,t)\mathcal{H}_{ep}(0,0)\rangle^*$.

Substituting Eqs.~(\ref{eq:Eph}), (\ref{eq:exponential_correlator}) and
~(\ref{eq:Phonon}) into Eq.~(\ref{eq:K}) and using Eq.~(\ref{eq:Gamma}) we obtain the phonon-assisted rate $\gamma_{ep}$.
While carrying out the spatial integration in Eq.~(\ref{eq:K}) we may neglect the position
dependence in the electronic correlation functions and replace them by $\Pi(K_j,1;x=0,t)$ (this is justified by the small value of the adiabatic parameter $s_T/v_F$). Doing so we
obtain
\begin{equation}\label{eq:Gammaep}
\gamma_{ep}=\tilde{g}^2_T\frac{\pi v_F}{\xi}\left(\frac{\pi T\xi}{v_F}\right)^{\frac{1}{2}}\left(\frac{\pi T\xi}{u}\right)^{\frac{K_{c+}}{2}}F_{ep}\left(\frac{\omega}{T},\frac{s_T k_F}{T}\right),
\end{equation}\label{eq:G}
where $\tilde{g}_T\equiv g_T/\sqrt{s_T v_F}$ is the dimensionless
coupling constant of e-ph interaction,
$F_{ep}(\frac{\omega}{T},\frac{s_T k_F}{T})$ is the scaling function
for the phonon-assisted decay rate. Introducing dimensionless
variables $X=x T/s_T$ and $Y=T t$ we can express it as
\begin{widetext}
\begin{eqnarray} \label{eq:F_ep}
F_{ep}\left(\frac{\omega}{T},\frac{s_T k_F}{T}\right)&=&\frac{T}{\omega}\int_0^\infty d Y\int_{-\infty}^{\infty} d X \left(e^{i\frac{\omega}{T}Y}-1\right)\cos{\left(\frac{2k_F s_T}{T} X \right)}\nonumber\\
&&\left\{\left[\frac{1}{\sinh^2{(\pi(X+Y_-))}}+\frac{1}{\sinh^2{(\pi(X-Y_-))}}\right]\left(\frac{1}{ \sinh{(\pi Y_-) \sinh (-\pi Y_-)}}\right)^{\frac{3+K_{c+}}{4}}
-c.c.\right\}.
\end{eqnarray}

Similarly to the Umklapp case, the real part of the above scaling
function can be expressed in a more convenient form,
\begin{equation} \label{eq:F_ep_real}
\Re F_{ep}\left(\frac{\omega}{T},\frac{s_T k_F}{T}\right)=\frac{T}{ \pi^2 \omega}\sinh \frac{\omega}{2 T} \int_{-\infty}^\infty d y\int_{-\infty}^{\infty} d x  \left[\frac{1}{\cosh^2{(x+y)}}+\frac{1}{\cosh^2{(x-y)}}\right] \frac{ \cos{ \frac{\omega \, y}{\pi T}} \,\cos{\frac{2k_F s_T\, x}{\pi T}}}{ (\cosh y)^{(3+K_{c+})/2}}.
\end{equation}
\end{widetext}

From the above equations it is clear that the phonon-assisted decay
rate is practically unaffected by doping as long as $T> 2k_F s_T$.
Because of the slow velocity $s_T$ of the phonon mode, with the
exception of very low temperatures, this condition is satisfied for
all practically accessible doping levels. In this regime
Eq.~(\ref{eq:F_ep_real}) simplifies to
\begin{eqnarray} \label{eq:F_ep_real_Beta}
\Re \!F_{ep}\!\left(\!\frac{\omega}{T}\!,\!0\!\right)\!&\!=\!&\!\frac{2\, T}{\pi^2 \omega}\sinh \frac{\omega}{2 T} \, 2^{\frac{3+K_{c+}}{2}} \times \nonumber \\
 &\!&\!B\!\left( \!\frac{3 \!+\!K_{c+}}{4} \!-\!\frac{i\omega}{2 \pi T}, \!\frac{3\! +\!K_{c+}}{4} \!+\!\frac{i\omega}{2 \pi T}\!\right)\!.
\end{eqnarray}
In the limit $\omega\ll T$, the scaling function goes to a constant,
$F_{ep}(0,0)=2^{\frac{3+K_{c+}}{2}} B( [3 +K_{c+}]/4,[3
+K_{c+}]/4)/\pi^2$.

Similarly to the case of Umklapp processes the phonon assisted decay
rate can be expressed in terms a renormalized form, which does not
involve the short distance cutoff. In this case the role of the low
energy scale for the electron-phonon interaction with the TA mode is
played by the twist Peierls gap~\cite{Chen}, $\Delta_{ep}\sim
\frac{v_F}{\xi} \, \tilde{g}_T^{4/(1-K_{c+})}$, which depends on the radius of the CNT as $\sim R^{-[1+2/(1-K_{c+})]}$ and in the limiting case $K_{c+}\to 0$ is on the order of a few tens of Kelvin for a $(5,5)$ armchair CNT. This gap is induced by
the Peierls instability of the armchair tube with respect to a twist
distortion~\cite{Figge,Suzuura2000,Chen}. Up to a numerical coefficient on
the order unity the phonon-assisted decay rate can be expressed in
terms of the twist Peierls gap as,
\begin{equation}\label{eq:gamma_ep_renormalized}
    \gamma_{ep} \sim \Delta_{ep} \left( \frac{T}{\Delta_{ep}}\right)^{\frac{1+K_{c+}}{2}} \! F_{ep}\left(\frac{\omega}{T},\frac{s_T k_F}{T}\right).
\end{equation}
In the dc limit and at zero doping, the Luttinger liquid effects manifest themselves by changing the power dependence of $\gamma_{ep}$ or resistivity from $\gamma_{ep}\sim T$ for non-interacting case~\cite{KaneandMele} to $\gamma_{ep}\sim T^\frac{1+K_{c+}}{2}$ in a Luttinger liquid at temperatures above the Peierls gap. Experiments on phonon-induced resistivity to date~\cite{KaneandMele,Park} have been conducted on CNTs placed on substrates rather than freely suspended CNTs, and exhibit linear temperature dependence of resistivity, consistent with the non-interacting electron approximation, $K_{c+}=1$.

\section{summary of the results \label{sec:summary}}

We studied the intrinsic plasmon decay rates in clean armchair
nanotubes. Plasmon oscillations decay into neutral electron modes.
Such processes are facilitated by electron-electron and
electron-phonon backscattering processes. In second order of
perturbation theory in the backscattering interactions the total decay
rate is given by the sum of partial decay rates due to each mechanism.
For undoped CNT, both electron-electron and electron-phonon rates
reveal power law dependence on the frequency of oscillations or the
temperature of the system with non-integer exponent, that depends on
the Luttinger liquid parameter $K_{c+}$.

The plasmon decay processes lead to temperature dependent corrections
to power dissipation in the device depicted in Fig.~\ref{fig:setup}.
Therefore they can be studied experimentally in a finite device by measuring the temperature
dependence of power dissipation. In the regime where the temperature length $L_T=v_F/T$ is shorter than the plasmon wavelength the temperature-dependent part of the dissipated power can be expressed in terms of the instrinsic plasmon decay rate $\gamma$ in homogeneous nanotube using Eqs.~(\ref{eq:powerreducedform}-\ref{eq:formfactor}) and (\ref{eq:Gamma}).
Furthermore, in this regime $\gamma$ can be related to the intrinsic ac conductivity
of carbon nanotubes via Eq.~(\ref{eq:conductivity}). Thus intrinsic plasmon decay rate in CNT can, in  principle, be extracted from measurements in finite size devices.

Among the electron-electron interactions resulting in plasmon decay
the most relevant ones are Umklapp processes. They correspond to
backscattering of pairs of electrons and lead to the decay rate that
scales as $\gamma_u\sim T^{2K_{c+}-1}$ at zero doping and becomes
exponentially small, $\sim \exp(-2k_Fv_F/T)$, at large doping, see
Eq.~(\ref{eq:gamma_u_main}). This rate can be expressed in terms of
the short distance cutoff and Umklapp coupling constants, see
Eq.~(\ref{eq:gamma_u_main}) or in terms of the cutoff-independent
Umklapp gap, as in Eq.~(\ref{eq:gamma_u_renormalized}).

In contrast to the Umklapp processes, the phonon-assisted decay rate depends on the doping level only on a very large scale $v_F k_F \sim T v_F/s_T \approx 0.6 \times 10^2 \, T$. With the exception of rather low temperatures the practically achievable doping levels lie below this scale. In this regime the phonon-assisted decay rate scales as
 $\gamma_{ep} \sim
T^{(1+K_{c+})/2}$, as given by Eq.~(\ref{eq:Gammaep}). It can also be expressed in
terms of a cutoff-independent energy scale $\Delta_{ep}$ associated
with electron-phonon interactions, see
Eq.~(\ref{eq:gamma_ep_renormalized}). This scale is given by the gap
in the electron spectrum induced by the twist Peierls instability.

If the plasmon frequency is smaller than the temperature, the ratio of
the phonon-assisted rate and the Umklapp one at zero doping is
$\gamma_{ep}/\gamma_u \sim \left( \frac{T^3 \Delta_{ep}}{\Delta_u^4}\right)^{\frac{1-K_{c+}}{2}}$. The phonon
assisted decay is dominant in the wide range of temperatures, $T >
(\Delta_u^4/\Delta_{ep})^{1/3}$. With the exception of the region of very low
temperatures it is dominated by electron-phonon scattering and scales
as $T^{(1+K_{c+})/2}$. This is consistent with the temperature dependence of phonon-induced resistivity~\cite{Kane1997}.

We are grateful to David Cobden, Dan Prober and Xiaodong Xu for useful discussions.
This work was supported by the DOE grants DE-FG02-07ER46452 (W.C. and A.V.A)
and DE-FG02-06ER46313 (E.G.M.) NSF DMR Grant No. 0906498 at Yale University (L.I.G.), and the Nanosciences Foundation at Grenoble, France. The hospitality of CEA Grenoble (France) and the Institute for Nuclear Theory at the University of Washington (Seattle) is greatly appreciated.

\appendix

\section{fusion rules \label{sec:fusion}}

Equation (\ref{eq:G^R}) can be derived by noting that both the Umklapp
and electron-phonon backscattering Hamiltonian densities have
exponential dependence of the charge field $\Phi_{c+}$, $\sim
(e^{ia_\eta \Phi_{c+}} + h.c.)$ and using the following identity that
holds for Gaussian averages,
\begin{widetext}
\begin{eqnarray}
&\langle \phi_{c+}(x)e^{ia_\eta\phi_{c+}(x_1)}e^{-ia_\eta\phi_{c+}(x_2)}\phi_{c+}(x')\rangle_0=\nonumber\\
&-(ia_\eta)^2\langle\phi_{c+}(x)\phi_{c+}(x_1)\rangle_0 \langle e^{ia_\eta\phi_{c+}(x_1)}e^{-ia_\eta\phi_{c+}(x_2)}\rangle_0 \langle\phi_{c+}(x_2)\phi_{c+}(x')\rangle_0 \nonumber\\
&+ (ia_\eta)^2\langle\phi_{c+}(x)\phi_{c+}(x_1)\rangle_0 \langle e^{ia_\eta\phi_{c+}(x_1)}e^{-ia_\eta\phi_{c+}(x_2)}\rangle_0 \langle\phi_{c+}(x_1)\phi_{c+}(x')\rangle_0 + \nonumber\\
& \text{same  terms with} \ x_1 \leftrightarrow  x_2 + \langle\phi_{c+}(x)\phi_{c+}(x')\rangle_0\langle e^{ia_\eta\phi_{c+}(x_1)}e^{-ia_\eta\phi_{c+}(x_2)}\rangle_0. \label{eq:fusion}
\end{eqnarray}
\end{widetext}
This formula can be established by expanding the exponentials in the
left hand side in the Taylor series and applying Wick's theorem to
each term. The resulting series sums to the expression in the right
hand side.

The last term in the right hand side corresponds to a disconnected
diagram and does not contribute to the correlation function.

\end{document}